\documentstyle[epsfig]{article}

\title{Gain Dependence of the Noise in the Single Electron Transistor}
\author{B. Starmark, Torsten  Henning, T. Claeson and P. Delsing\\
Department of Microelectronics and Nanoscience, G\"oteborg University\\
and Chalmers University of Technology, S-412 96 G\"oteborg, Sweden
\and
A. N. Korotkov\\
Nuclear Physics Institute, Moscow State University,
119899 GSP, Russia}

\begin{document}

\maketitle

\begin{abstract}
An extensive investigation of low frequency noise in single electron
transistors as a function of gain is presented.
Comparing the output noise with gain for a large number of bias points,
it is found that the noise is dominated by external charge noise.
For low gains we find an additional noise contribution which is compared to
a model
including resistance fluctuations. We conclude that this excess noise
is not only due to resistance fluctuations.
For one sample, we find a record low minimum charge noise of $q_{\rm
n}\approx 9\cdot 10^{-6}$\,$\rm e/\sqrt{\rm Hz}$ in the superconducting state and
$q_{\rm n} \approx 16\cdot 10^{-6}$\,$\rm e/\sqrt{\rm Hz}$ in the normal 
state at a frequency of 4.4~kHz.
\end{abstract}

\section{Introduction}
With the introduction of the Single Electron Transistor (SET) one
decade ago, it became possible to directly measure changes in charge
quantities below that of an electron
\cite{Likharev,Fulton}. Based on the Coulomb blockade, the
device has been shown to be the most sensitive electrometer existing today.
The sensitivity of the SET is predicted to be limited by the shot noise
\cite{Schottky}
generated when electrons tunnel across the tunnel
barriers \cite{Korotkov}. Shot noise was observed in a two
junction structure (without gate) \cite{Birk}.
In most experiments involving SETs, the noise
for low frequencies has been dominated by the device itself, whereas
external sources set the noise limit for frequencies above the kHz
regime.

Several experimental studies of low frequency noise of various SET
configurations have been performed \cite{Kuzmin}-\cite{PTB}. Below 1
kHz, $1/f$ noise is observed in all SETs regardless of mode of
operation \cite{Zimmerli1}-\cite{Zorin2}.
The input equivalent charge noise at 10 Hz in all these experiments
is of the order of $10^{-3}$ to $10^{-5}$ $\rm e/\sqrt{Hz}$, with $2.5
\cdot 10^{-5}$ $\rm e/\sqrt{Hz}$ recently reported as the lowest figure
\cite{Krupenin}.
Deviations from an $1/f$ spectrum are often observed, usually
in combination with telegraph noise \cite{Zimmerli1,Song}.
The source of the latter is believed to be random excitations of a single
charge trap. Theoretically, the random
trapping process of a single trap shows a Debye-Lorentzian power
spectrum \cite{Machlup} which is also observed experimentally
\cite{Verbrugh,Bouchiat,Zimmerman}. It has been shown that an ensemble
of traps produces a $1/f$ noise spectrum, see {\it e.g.} \cite{Rogers}.

There are at least three possible locations of these fluctuators: the
tunnel junction
dielectric, the substrate on which the device is fabricated, and the
oxide layer covering the device.
The role of the substrate has been examined in at least two
sets of experiments \cite{Bouchiat,PTB}. Those experiments did not show
a strong dependence of the noise on the substrate material.
The barrier dielectric has been proposed as the location of
charge traps \cite{Verbrugh,Zorin1,Song,Krupenin}. The role of the
surface oxide of the island has not yet been investigated.

Several groups have found that the noise at the output of the SET
varies with the gain of the SET and that the maximum
noise is found at the bias point with maximum gain
\cite{Verbrugh}-\cite{Tavkhelidze}. This indicates that
the noise source acts at the input of the device, {\it i.e.} as an external
fluctuating charge. However, a detailed comparison of the noise to the gain
has not been done.

In this Letter, we report the low frequency current noise of one\\
${\rm Al/Al_2O_3/Al/Al_2O_3/Al}$ SET and one ${\rm
Nb/Al_2O_3/Al/Al_2O_3/Nb}$ SET and make a detailed comparison with the
gain. (Hereafter, we
will refer to the two SETs as the Al SET and the Nb SET). For the Al SET,
we find that the
noise follows the gain in such a manner that the SET is dominated by input
noise for almost all values of bias and gate voltage. For the Nb SET
however, we find a contribution from other sources when analysing the noise
in the points where the gain is low.

\section{Experimental Techniques}

% Fabrication
The samples were fabricated on oxidised Si substrates using
electron beam lithography and the standard double-angle evaporation
technique \cite{Dolan}.

The resistance of the Al SET directly after fabrication was
$R_{\rm T}=R_{1}+R_{2}\approx 0.8$\, k$\rm \Omega$, which after a storage for six
months, had increased to $R_{\rm T}
\approx 45$\, k$\rm \Omega$. The Nb SET had a resistance of $R_{\rm T}\approx 170$\,
k$\rm \Omega$.

% Electronics
We used a symmetric, current sensitive amplifier which voltage
biased the SET \cite{Starmark}. To
optimize the preamplifier noise performance low noise operational
amplifiers with low $1/f$ noise were used. Furthermore, the
bias (feedback) resistors were chosen to $R_{\rm F} = 10$\,M$\rm \Omega$ to lower the
amplifier noise floor at low frequencies.

% Measurement Setup
The SETs were attached to the mixing chamber of a dilution refrigerator
which was cooled to a
temperature below 30\,mK. All measurement leads were filtered
with 0.5\,m Thermocoax \cite{Zorin3} followed by capacitors to ground.
The total line capacitance was $C_{\rm l}$=1\,nF.

% AC Performance
Evaluating the frequency performance, we found a gain bandwidth of the SET
setup of 7.5\,kHz while the noise bandwidth was about 300 Hz (without any SET
connected).
Both these figures were set by the line capacitance and the
preamplifier \cite{Starmark}.

% Machinery
The noise spectra were recorded by a HP 35665A Dynamic Signal
Analyzer, which performs real-time FFT analysis of the input signal.
The frequency range from 1 to $10^5$ Hz was divided into four subranges to
increase the
resolution. The time to acquire all noise data for one bias point was
5 min.

\section{Results and Discussion}

\begin{figure}[t]
\noindent
\begin{minipage}[c]{0.96\textwidth}
\centering
\epsfig{file=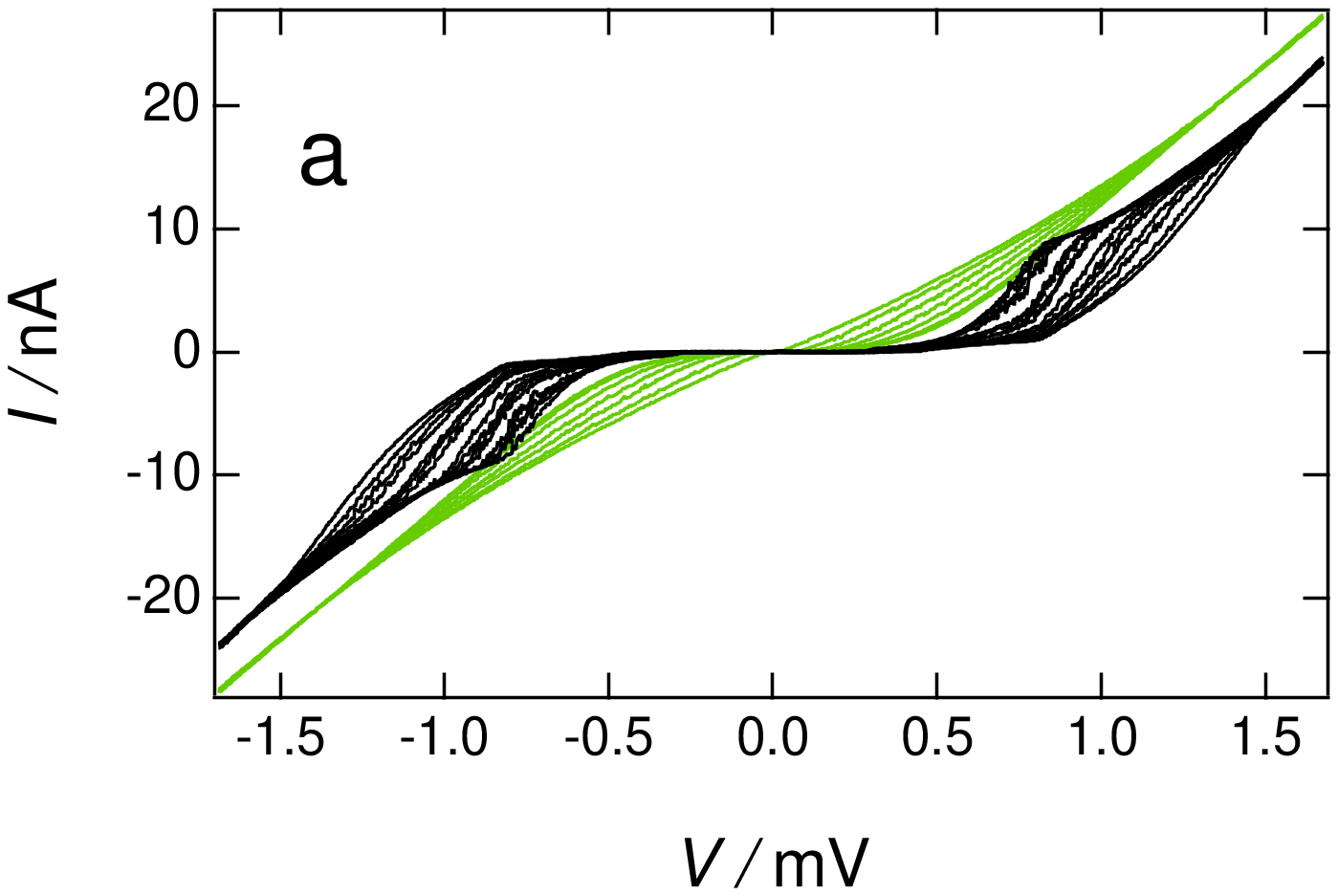,width=0.96\textwidth}
\end{minipage}
\nobreak
\begin{minipage}[c]{0.96\textwidth}
\centering
\epsfig{file=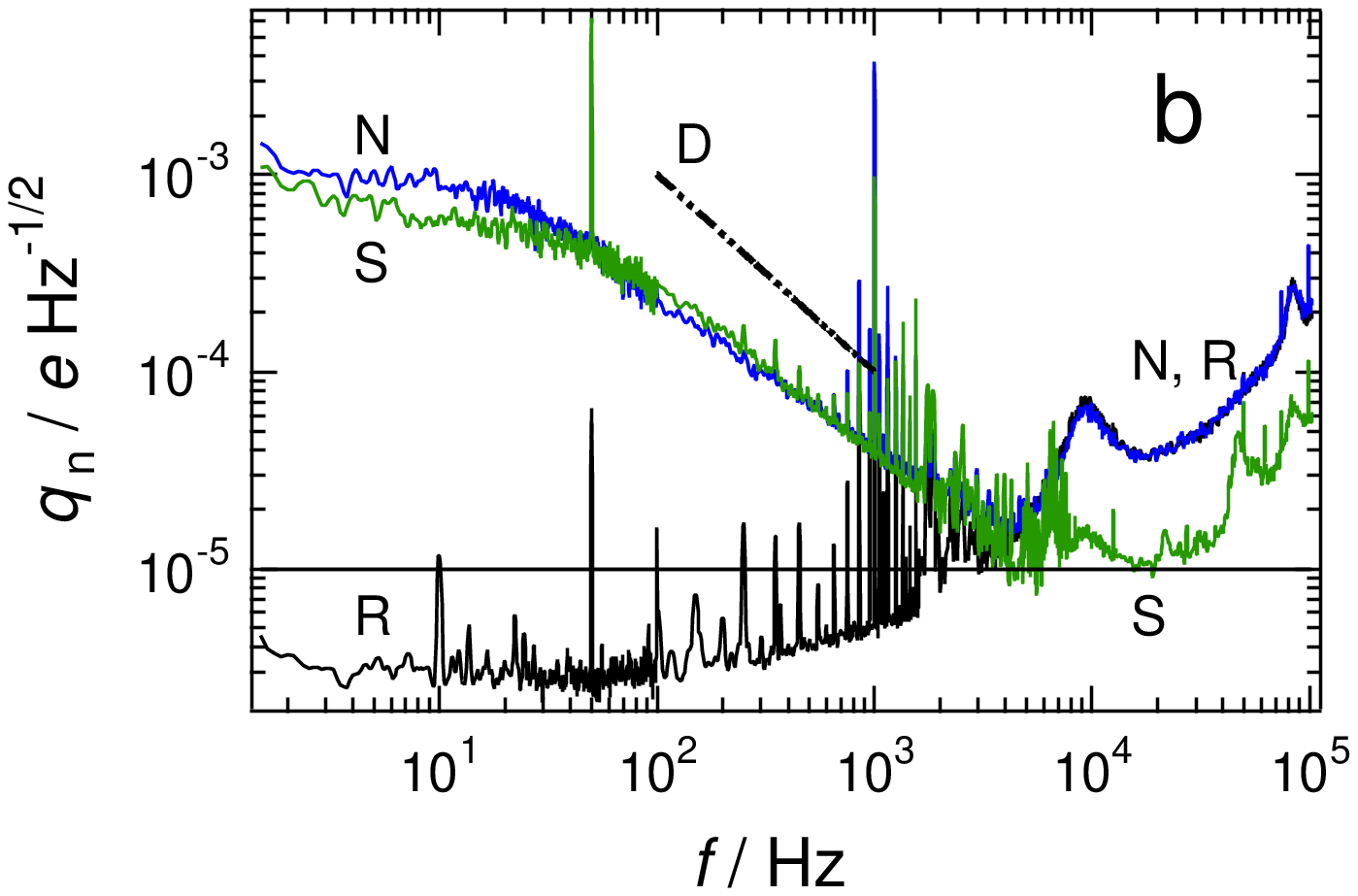,width=0.96\textwidth}
\end{minipage}

\caption{\label{IV-char+spectra}
a) Current-voltage characteristics for the Al SET in the normal and
superconducting
state for several different gate charges. The curves were measured at
$T=30$\,mK.
b) Noise spectra for the system for the normal (N) and superconducting
(S) state
and with no SET connected (R). The noise has been input referred to a
charge noise (see text). Curve D shows a $1/f^2$ slope.}
\end{figure}

% IV-curves: Rt, Csum and Rout
Two samples were tested. The current-voltage
$IV$ characteristics
for the Al SET are shown in Fig.~\ref{IV-char+spectra}a, both for
the normal and the superconducting state.
A total island capacitance of $C_\Sigma=0.19$\,fF
was deduced from the IV-curves. The Nb SET had $C_\Sigma=0.48$\,fF.
The output impedance, $r_{\rm o} = (\partial
I/\partial V )$, was calculated from the $IV$
curves.
In the superconducting state $r_{\rm o}$ was always above
$20$\,k$\rm \Omega$ while in the normal
state, $r_{\rm o}$ was on the order of, or above $R_{\rm T}$ for both SETs.
The gate coupling capacitances were $C_{\rm g}\approx4.8$\,aF and $C_{\rm
g}\approx0.3$\,aF
for the Al and the Nb SETs, respectively.

Due to its low resistance, the AL SET had a very high maximum gain of
$\partial I/\partial Q_{\rm g}=12$\,nA/e
and $\partial I/\partial Q_{\rm g}=34$\,nA/e
in the normal and superconducting states, respectively. The Nb SET had
$\partial
I/\partial Q_{\rm g}=1.8$\,nA/e
and $\partial I/\partial Q_{\rm g}=3.8$\,nA/e in the normal and superconducting
states.
The gain increase in the superconducting state is in accordance with
earlier observations \cite{Hergenrother,Song}.

% Noise Spectra
Noise spectra for the normal and superconducting states of the Al SET are shown
in Fig.~\ref{IV-char+spectra}b. Each spectrum
has been referred to the input of the SET by dividing by the frequency
dependent gain.
The spectra N and S were measured at the bias points which gave maximum gain.
For reference, a spectrum with no SET connected, R,
is also shown and is divided by the same gain as in the normal
state to obtain the input referred noise floor set by the amplifier.
Minimum charge noises of $q_{\rm n}\approx 9\cdot 10^{-6}$\,$\rm e/\sqrt{\rm Hz}$
and $16\cdot 10^{-6}$\,$\rm e/\sqrt{\rm Hz}$ at a
frequency of 4.4\,kHz was found in the superconducting and normal
states,
respectively. The limit is set by the preamplifier and mechanical resonances
within the cryostat. These numbers are, to our knowledge, the lowest
values reported for any SET. The noise at 10
Hz was $7\cdot 10^{-4}$\,$\rm e/\sqrt{\rm Hz}$ and
$9\cdot 10^{-4}$\,$\rm e/\sqrt{\rm Hz}$ for the superconducting
and normal states, respectively.
A cross-over from input dominated to output dominated
noise can be seen as the frequency increases.
Below 1 kHz, the input referred noise is actually the same in
both normal and superconducting states, indicating that the noise source
acts as an apparent charge noise, and thus is independent of gain.
Above 1 kHz the noise is dominated by sources acting at the output of
the SET. When referred to the input, this noise appears
as a lower equivalent charge noise in the superconducting state as compared
to the
normal state, due to the higher gain in the superconducting state.
\begin{figure}[t]
\noindent
\begin{minipage}[c]{0.96\textwidth}
\centering
\epsfig{file=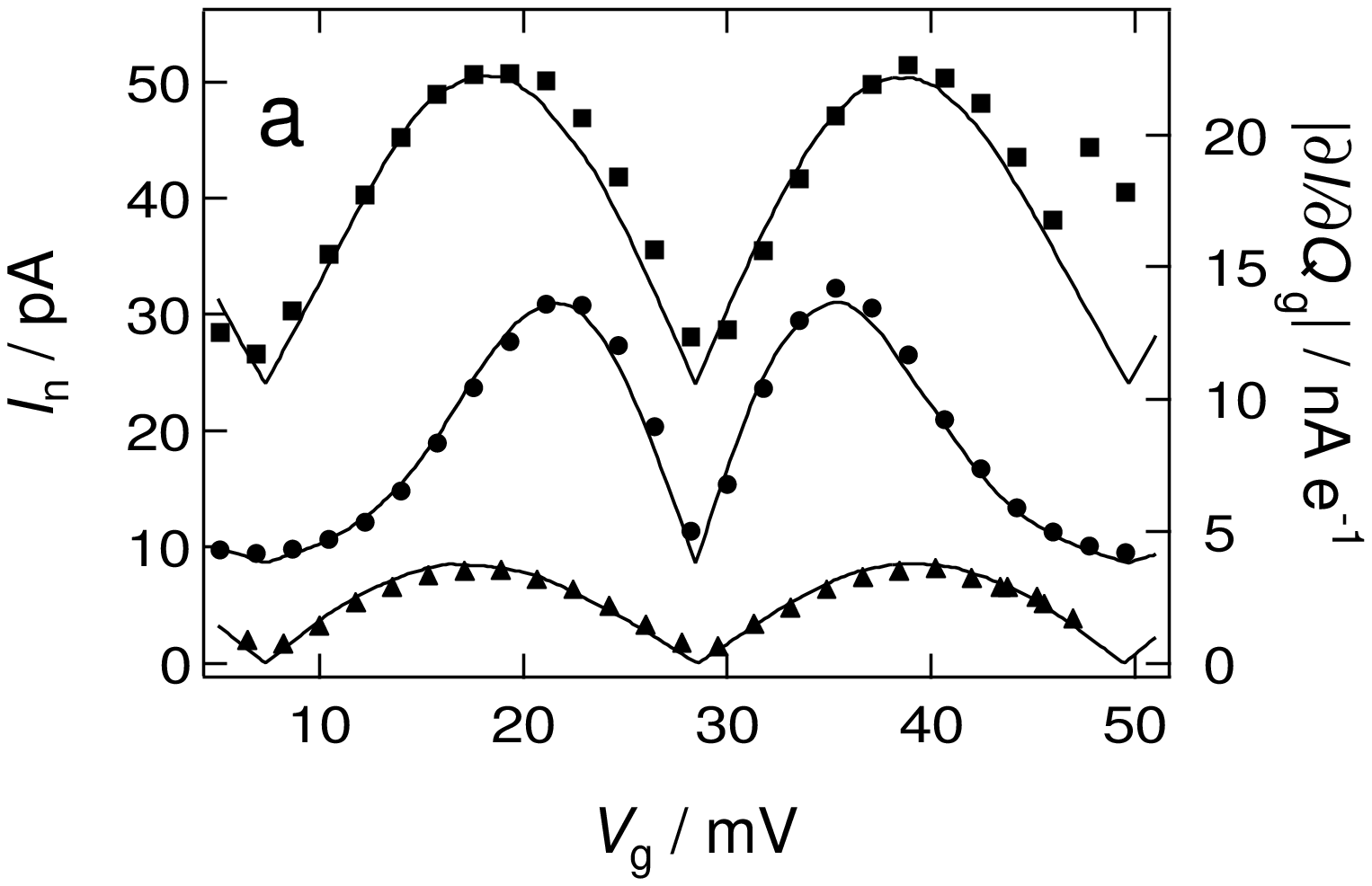,width=0.96\textwidth}
\end{minipage}
\nobreak
\begin{minipage}[c]{0.96\textwidth}
\centering
\epsfig{file=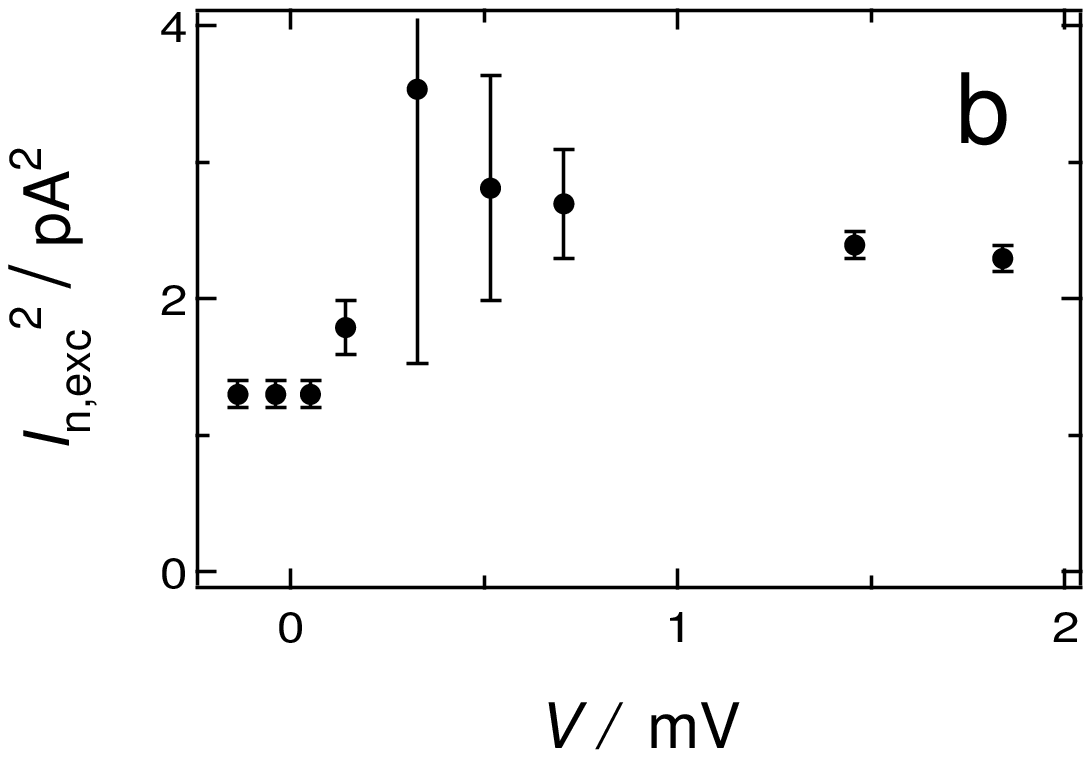,width=0.96\textwidth}
\end{minipage}

\caption{\label{noisegain}%
a) Gain (curves) and integrated output noise (symbols) vs gate bias for the
Al SET. The bias voltage is
$V=0.25$\,mV (middle curve), $V=0.43$\,mV (upper curve) and
$V=1.04$\,mV
(lower curve). The curves have been offset vertically for clarity.
b) excess noise as a function of bias voltage for the Nb SET.}
\end{figure}

% Noise and gain
We now turn our attention to the noise below 1 kHz. To
determine the origin of the noise, we measured
noise for 130 bias and gate voltages and compared it with the
measured gain of the SET. All points were taken in the normal state
at a temperature of $T\approx 30$\, mK. The output current
noise and the gain versus gate bias for the Al SET are shown in
fig.~\ref{noisegain}a for three
bias voltages. To reduce fluctuations, the noise was integrated in the
band 51 to 99 Hz.
By using the gate charge
offset as the only fitting parameter, we were able to get an excellent
fit of the noise to the gain.
This result clearly shows that the noise source acts at the
input of the SET. It is supposedly due to the motion
of background charges somewhere in the vicinity of the
SET-transistor. For the points of large gain we can deduce the
spectral density of the charge fluctuations, $S_{Q_{\rm g}} (f)$. The Nb SET also
showed a gain dependent output noise, but with more spread in the data.

In bias points with low
gain (high bias or $Q_{\rm g}\approx n\rm e/2$)
it is possible to examine other noise contributions. For each $V$, we
define the excess noise
current as

\begin{equation}
\centering
{{I_{\rm n,exc}^2\left( {V} \right)=\left( {\int_{51\ {\rm Hz}}^{99\ {\rm
Hz}} {S_{I,{\rm min}}df-{{G_{\rm min}^2}
 \over {G_{\rm max}^2}} \int_{51\ {\rm Hz}}^{99\ {\rm Hz}} {S_{I,{\rm
max}}df}}} \right)} \mathord{\left/ {\vphantom {{I_{\rm n,exc}
^2\left( {V} \right)=\left( {\int_{51\ {\rm Hz}}^{99\ {\rm Hz}}
{S_{I,{\rm min}}df-{{G_{\rm min}^2} \over {G_{\rm max}^2}}
\int_{51\ {\rm Hz}}^{99\ {\rm Hz}} {S_{I,{\rm max}}df}}} \right)} {\left(
{1-{{G_{\rm min}^2} \over {G_{\rm max}^2}}} \right)}}}
\right. \kern-\nulldelimiterspace} {\left( {1-{{G_{\rm min}^2} \over
{G_{\rm max}^2}}} \right)}}
\label{excess_noise_def}
\end{equation}

where ${S_{I,\min}}$, ${G_{\min}}$ and ${S_{I,\max}}$, ${G_{\max}}$
 are the measured current power spectra and gains for each $V$ at minimum and
 maximum gain respectively. This quantity is zero when the gain dependent
 noise is the only contribution.

The Al SET showed excess noise only at the highest bias
 point, $2.6\pm 0.97\ \rm{pA^2 / Hz}$, whereas the Nb SET showed excess noise
 at several bias points.
To gain more knowledge about the noise contributions
we next discuss the different components of the measured noise.

Except for charge noise sources somewhere in the vicinity of the transistor,
current noise can also be induced by fluctuations in the tunnel barrier
resistance. As we will see, these two contributions can generally not be
separated.
There are also contributions from shot noise and amplifier noise.
In our case, the amplifier noise is dominated by the thermal noise of the
two feedback
resistors and the noise generated by the input equivalent voltage noise,
$e_{\rm n}$, of the amplifier.
For low temperatures, the shot noise is given by $S_I=aeI$, where $1\le a\le 2$,
with $a=2$ for strongly correlated tunneling and $a=1$ for uncorrelated
tunneling
\cite {Korotkov}.
The total measured current noise in the system can be modelled by

\begin{eqnarray}
\centering
S_{I_{\rm m}}(f) = 2k_{\rm B} \frac{T_{\rm F}}{R_{\rm F}} +\frac {e_{\rm n}^2(f)}{2 r_{\rm o}^2}
 + aeI + S_{I_{\rm Q,R}}(f)
\label{totalnoises}
\end{eqnarray}

where $T_{F}$ is the temperature of the feedback resistors and
$S_{I_{\rm Q,R}}(f)$ is the combined spectral density due to charge and
resistance fluctuations.
The first two terms represent amplifier noise and are frequency independent
in the range of interest  (51-99 Hz).
The third term is the shot noise of the SET which also is independent of
frequency. These terms are of order $\left( {30\ \rm fA} \right)^2/\rm Hz$
for the worst case
with lowest $r_{\rm o}$ and high I.
Thus, the sum of these three terms varies only slightly
between $\left( {30\ \rm fA} \right)^2/\rm Hz$ and $\left( {50\ \rm fA}
\right)^2/\rm Hz$ for all bias
points.
To evaluate the last term let us start with a model in which the only
fluctuating parameter is the background charge $Q_{\rm g}$. Assuming small
variations we can write

\begin{equation}
\delta I = \frac{\partial I}{\partial Q_{\rm g}} \delta Q_{\rm g} +
\frac{1}{2} \, \frac{\partial^2 I}{\partial Q_{\rm g}^2} (\delta Q_{\rm g})^2 + \dots
\label{I(Q)}
\end{equation}

If only the first (linear) term is taken into account, then obviously
$S_{I_{\rm Q}}(f)=(\partial I/\partial Q_{\rm g})^2 S_{Q_{\rm g}} (f)$ where 
$S_{Q_{\rm g}}$ and $S_{I_{\rm Q}}$ are the background
charge spectral densityand the charge induced current spectral density.  As seen in fig.\,\ref{noisegain}a 
we are close to this situation in the experiment.

Close to the operation points for which
$\partial I/\partial Q_{\rm g} =0$, the contribution from the quadratic term in
eq.\,(\ref{I(Q)}) becomes important. Assuming Gaussian noise we get

\begin{equation}
S_{I_{\rm Q}}(f)\approx \left( \left( \frac {\partial I}{\partial Q_{\rm g}} \right)^2 +
\frac{\alpha}{4} \left( e \frac{\partial^2 I}{\partial Q_{\rm g}^2} \right)^2\right)
S_{Q_{\rm g}} (f)
\label{SI(Q)}
\end{equation}

where $\alpha (f) = [\int_{-\infty}^{+\infty}
S_{Q_{\rm g}}(f')S_{Q_{\rm g}}(f-f')df']/e^2S_{Q_{\rm g}}(f)$.
For the SETs, we find $\alpha(f) \sim 10^{-4}$.
This is smaller than the two orders of magnitude of dynamic range we have
in the noise measurement and we can thus neglect the second term.

Now let us consider a different model in which $Q_{\rm g}$
does not fluctuate, and the only fluctuating parameter is the tunnel
resistance $R_1$ of the first junction. (The fluctuations of the tunnel
junction resistance have been extensively studied in large area
junctions, see e.g., Refs.\ \cite{Rogers}.)  For simplicity let us
limit ourselves to the linear term of the series expansion, so that
$S_{IR_1}(f)= (\partial I/\partial R_1)^2 S_{R_1}(f)$, where 
$\partial I/\partial R_1$ can be calculated \cite{Korotkov}.
Note that $S_{IR_1}$ is asymmetric around $Q_{\rm g}=\rm e/2$ as a function of $Q_{\rm g}$ 
for $V \le {e \over {2C_\Sigma }}$ \cite{comment},
while it becomes independent of $Q_{g}$
and proportional to $V^2$ for large $V$.
Furthermore, the fluctuations of $R_1$ can in principle
change with $V$, $Q_{\rm g}$, and $T$ (however, a strong
dependence on $V$ and $Q_{\rm g}$ is unlikely).

On the other hand the current noise which is due to $Q_{\rm g}$
fluctuations decreases for sufficiently large $V$ because of
the decrease in $|\partial I/\partial Q_{\rm g}|$ and is symmetric
around $Q_{\rm g}=\rm e/2$ for a symmetric SET-transistor.
Therefore, the bias dependence and this asymmetry could be used  
to distinguish the charge fluctuations from the resistance fluctuations.

In general, the noise can also be caused by simultaneous fluctuations
of $Q_{\rm g}$, $R_1$, and $R_2$. If they are uncorrelated, the corresponding
current spectral densities should simply add. However, fluctuators inside
the tunnel barrier, which have been suggested as the source of the $1/f$
noise by
several authors \cite {Verbrugh,Zorin1,Song,Krupenin}, can be responsible for
both resistance and charge fluctuations. Then, these contributions
are correlated and we arrive at the following expression

\begin{eqnarray}
S_{I_{\rm Q,R}}(f) = \left( \frac{\partial I}{\partial Q_{\rm g}}\right) ^2 S_{Q_{\rm g}} (f) +
\left( \frac{\partial I}{\partial R_1}\right) ^2 S_{R_{\rm 1}} (f) +
\left( \frac{\partial I}{\partial R_2}\right) ^2 S_{R_{\rm 2}} (f)
\nonumber \\
+ K_1 \frac{\partial I}{\partial Q_{\rm g}} \, \frac{\partial I}{\partial R_1}
\sqrt{S_{Q_{\rm g}} (f) S_{R_{\rm 1}} (f)}
+ K_2 \frac{\partial I}{\partial Q_{\rm g}} \, \frac{\partial I}{\partial R_2}
\sqrt{S_{Q_{\rm g}} (f) S_{R_{\rm 2}} (f)}
\label{allnoises}
\end{eqnarray}

$K_i$ is the dimensionless correlation coefficient between
$Q_{\rm g}$ and $R_{\rm i}$ fluctuations, $|K_i|\le 1$.

The first term in eq.\,(\ref{allnoises}) describes the dominating gain
dependent noise, while the other terms contribute to the excess noise.
We can now compare the bias dependence of the excess noise measured in the Nb SET
to that of eq.\,(\ref{allnoises}).
From the integrated noise spectra we calculate
the measured excess current noise $I_{\rm{n,exc}}$, according to
eq.\,(\ref{excess_noise_def})
and plot it versus $V$ in fig.\ref{noisegain}b. $I_{\rm{n,exc}}$ seems to
increase with bias voltage, but there is no quadratic dependence
which would be expected from eq.\,(\ref{allnoises}). It thus seems likely
that the excess noise is {\it not} due to resistance fluctuations, but has a
different origin. One possible explanation is that the increasing current
heats the SET and generates more noise.
Furthermore, at the highest bias point we can set an 
upper level for the resistance fluctuation in the frequency range from 
51-99 Hz, $\delta R_{51-99}$. Assuming a symmetric SET ($R_{\rm 1}=R_{\rm 2}$
and $S_{R_{\rm 1}}=S_{R_{\rm 2}}$) we get $\delta R_{51-99}<30$\,$\Omega_{\rm RMS}$ for the
Nb SET and $\delta R_{51-99}<5$\,$\Omega_{\rm RMS}$ for the Al SET.

In conclusion we have measured the low frequency noise of two single
electron transistors. In both transistors, the noise at the output closely
followed the gain. This shows that low frequency noise in the SET is 
mainly due to external charge noise. When the gain was low, we 
observed an excess noise in the Nb SET for all bias voltages and in the 
Al SET for the highest bias voltage. From the bias dependence 
of the excess noise in the Nb SET we conclude that the main source of the 
excess noise is not resistance fluctuations. We also set an upper 
limit for the resistance fluctuations. The Al SET had a very high gain and 
showed a minimum charge
noise $q_{\rm n}\approx 9\cdot 10^{-6}$\,$\rm e/\sqrt{\rm Hz}$
and $16\cdot 10^{-6}$\,$\rm e/\sqrt{\rm Hz}$ in the
superconducting and normal state, respectively.

\section{Acknowledgements}
We gratefully acknowledge discussions with K.K. Likharev and A.B. Zorin.
The samples were fabricated in the Swedish Nanometer
Laboratory. This work was sponsored by the Swedish SSF and
NFR, by the ESPRIT project CHARGE and by the Japanese NEDO.


\begin{thebibliography}{99}
\bibitem{Likharev}Likharev K. K., {\it IEEE Trans. Mag.} {\bf 23},
1142 (1987)

\bibitem{Fulton} Fulton T. A. and Dolan G. J., {\it Phys. Rev.
Lett.} {\bf 59}, 109
(1987)

\bibitem{Schottky} Schottky W., {\it Ann. Phys.} {\bf 57}, 541 (1918)

\bibitem{Korotkov} Korotkov A. N., Averin D. V., Likharev K. K. and
 Vasenko S. A.
in {\it Single Electron Tunneling and Mesoscopic Devices}, edited by H.
Koch (Springer, New York, 1992) p. 45

\bibitem{Birk} Birk H., de Jong M. J. M. and 
Schoenenberger C., {\it Phys. Rev. Lett.} {\bf 20}, 1610 (1995)

\bibitem{Kuzmin} Kuzmin L. S., Delsing P., Claeson T. and 
Likharev K. K., {\it Phys. Rev. Lett.} {\bf 62}, 2539 (1989)

\bibitem{Zimmerli1} Zimmerli G., Eiles T. M., Kautz R L. and 
Martinis J.M., {\it Appl. Phys. Lett.} {\bf 61}, 237 (1992)

\bibitem{Verbrugh} Verbrugh  S. M., Benhamadi M. L.,
Visscher E. H. and Mooij J. E., {\it J. Appl. Phys.} {\bf 78}
(1995), 2830

\bibitem{Zorin1} Zorin A. B., Ahlers F.-J. , Niemeyer J., Weimann T.,
Wolf H., Krupenin V. A. and Lotkhov S. V., {\it Phys. Rev. B} {\bf
53}, 13682 (1996)

\bibitem{Starmark} Starmark B., Delsing P., Haviland D. B. and 
Claeson T., in {\it Ext. Abstr. 6th International Superconductive Electronics
Conference}, pp. 391-393, (Berlin, 1997)

\bibitem{Tavkhelidze} Tavkhelidze A. N. and Mygind J., {\it J. Appl.
Phys.}
{\bf 83}, 310 (1998)

\bibitem{Bouchiat} Bouchiat V., Ph.D. Thesis, Paris 6 (1997), pp. 202

\bibitem{Zimmerman} Zimmerman N. M., Cobb J. L. and Clark A.
F., {\it Phys. Rev. B} {\bf 56}, 7675 (1997)

\bibitem{Hergenrother} Hergenrother J. M., Tuominen M. T., Tighe T. S. and
 Tinkham M., {\it IEEE Trans. Appl. Supercond.} {\bf 3}, 1980 (1993)

\bibitem{Song} Song D., Amar A., Lobb C. J. and Wellstood F. C.,
 {\it IEEE Trans. Appl. Supercond.} {\bf 5}, 3085 (1995)

\bibitem{Zorin2} Zorin A. B., Pashkin Y. A., Krupenin V. A. and 
Scherer H., in {\it Ext. Abstr. 6th International Superconductive Electronics
Conference}, pp. 394-396, (Berlin, 1997)

\bibitem{Krupenin} Krupenin V. A., Presnov D. E., Scherer H.,
Zorin A. B. and  Niemeyer J., cond-mat/9804197.

\bibitem{PTB} Niemeyer J. et. al., private communication

\bibitem{Machlup} Machlup S., {\it J. Appl. Phys.} {\bf 25}, 341 (1954)

\bibitem{Dolan} Dolan G. J., {\it Appl. Phys. Lett.} {\bf 31}, 337 (1977)

\bibitem{Zorin3} Zorin A. B., {\it Rev. Sci. Instrum.} {\bf 66}, 4296
(1995)

\bibitem{Rogers} Rogers C. T. and Buhrman R. A., {\it Phys. Rev.
Lett.}
{\bf 53}, 1272 (1984); {\it Phys. Rev. Lett.} {\bf 55}, 859 (1985)
 Savo B., Wellstood F. C., and Clarke J.,
{\it Appl. Phys. Lett.} {\bf 50}, 1757 (1987).

\bibitem{comment}
The noise is larger when the current is determined
by the fluctuating junction resistance to a greater extent than by the
other junction.

\end{thebibliography}
\end{document}